\documentclass[twocolumn,twoside,slac_two]{revtex4}
\usepackage{graphicx}
\usepackage{fancyhdr}

\usepackage[utf8]{inputenc}

\begin{document}

\title{A new interpretation of Giant radio pulses from the Crab pulsar}

%

\author{N. Lewandowska, C. Wendel, D. Elsässer, K. Mannheim}
\affiliation{University of Würzburg, Emil-Fischer Strasse 31,  97074 Würzburg , Germany}
\author{V. Kondratiev}
\affiliation{ASTRON, Netherlands Institute for Radio Astronomy}

\begin{abstract}
The Crab pulsar experienced a major flare in 2010 as observed by Fermi LAT. Observations by the Hubble Space Telescope indicate that the flare was accompanied by a structural change 
in the anvil region of the Crab Nebula. In the framework of a photometric analysis we reconstruct the energetics of this event. Reconnection zones near the light cylinder are expected 
to release energy by accelerating beams of electrons, leading to flares of varying amplitude. In this case the major flare would have reduced the magnetic energy stored in the 
reconnection zones, and would thus have had an impact on the properties of the giant radio flares presumably originating from these regions. We test this scenario by observing giant
radio pulses with the Westerbork Synthesis Radio Telescope.
\end{abstract}

\maketitle

\thispagestyle{fancy}


\section{Introduction}
Providing apparently the central driving mechanism of the Crab Nebula makes the Crab pulsar a unique object among other pulsars.
Being detected for the first time during balloon measurements of the Crab Nebula at hard X-rays (\cite{fishman_1969}), it is the only pulsar known so far with a pulsed emission profile which prevails throughout the whole electromagnetic spectrum from radio to GeV energies (\cite{mof}).
The profile contains several components known as precursor, main pulse (P1), interpulse (P2) and the high frequency components HFC 1, HFC 2 among others.
Although the precursor and both high frequency components are visible just at a certain wavelength range, P1 and P2 prevail from radio to gamma wavelengths.\\
Apart from its regular pulse structure the Crab pulsar is also a powerful emitter of giant radio pulses (GPs). These pulses distinguish themselves by several characteristics which 
will be briefly summarized here.  
Since the detection of the Crab pulsar at radio wavelengths by its GPs (\cite{staelin}), several properties have been observed like flux densities higher by at least thousand times than regular pulses.
Furthermore their widths are smaller in contrast with regular pulses. They vary from several microseconds down to 2 nanoseconds (\cite{hankins_2003}) while the shortest ones have 
been observed to have the highest flux densities. Being observed at a frequency range from 23 MHz (\cite{popov_2006}) till 15.1 GHz (\cite{hankins_2000}, \cite{jessner_2010})
reveals GPs apparently as a broadband phenomenon.
They have been observed mainly at the phases of P1 and P2 overlapped with regular pulses although they are apparently non-periodical (one GPs occuring every 0.803 seconds according to \cite{karuppusamy_2010}), 
However, there seem to be differences in the phases at which they occur since they were detected at the phases of HFC 1 and HFC 2 (\cite{hankins_2000},\cite{jessner_2005}), 
but not at the precursor for instance. Deducing from this they are apparently phase-bounded.\\
The characteristics of GPs have been studies largely by \cite{lundgren_1995} who observed Crab radio GPs also at $\gamma$ wavelengths. According to this study Crab GPs represent
single events and follow Poisson statistics. A comparison of the arrival times of both regular pulses and GPs gives indicates no difference between the arrival times of both 
which on the other hand suggests the same location of formation.\\
Simultaneous radio and optical observations of Crab GPs reveal by \cite{shearer_2003} an increase of the optical flux during occuring radio GPs by 3 $\%$. Hence the emission
mechanism which causes GPs is apparently non-coherent since it emits throughout different parts of the electromagnetic spectrum.\\ 
Differences between GPs occuring at the phases of P1 and P2 were discovered by \cite{hankins_eilek_2007} resulting from observations with the Arecibo radio telescope above 4 GHz.
They determined different dynamic spectra for GPs occuring at P1 in contrast with the ones detected at the phase of P2. While Giant main pulses (GMPs) consist in their substructure
of narrow-band nanopulses, Giant interpulses (GIPs) reveal narrow emission bands of microsecond duration. These results indicate probable different emission mechanisms for the
main and the interpulses and question current pulsar emission theories.\\
Theoretical aspects of radio GPs have been broadely discussed (\cite{mikhailovskii_1985}, \cite{weatherall_2001}, \cite{hankins_2003}, \cite{petrova_2004}).
The current only model basing on observational data is the Lyutikov model (\cite{lyutikov_2007}) which can reproduce the emission bands of the GIPs at frequencies above 4 GHz. 
Whereas regular pulses are thought to develop on open magnetic field lines, the Lyutikov model emanates from a higher particle density in contrast with the Goldreich-Julian standard model.
GPs are produced near the last closed magnetic field line via magnetic reconnection events through which a high energy Lorentz beam is produced. The latter  
moves along the closed field line and dissipates via curvature radiation. Thus it also predicts the occurence of $\gamma$-ray emission during the emission of radio GPs.\\
Although several other pulsars apart from the Crab have been found to emit GPs (\cite{slowikowska_2007}), a uniform emission mechanism for radio GPs has not been found yet.\\
Searching for a possible origin of this rather exotic form of pulsar emission, we examined the optical emission resulting from the Crab flare 2010 
with a photometrical analysis of three exposures made by the Hubble Space Telescope (HST).
The central motivation for this analysis was an estimation of the synchrotron power of the so called anvil region located 
approximately 5 arcseconds from the pulsar (see Figure~\ref{reference_stars}) in which an increase of brightness was detected after the flare detected by AGILE in September 2010.\\
If we assume that the Crab Nebula is powered solely by the pulsar, it is interesting to ask if the latter also contributes energetically to the reappearing flares 
detected from the Crab Nebula. We test this idea with GPs observed after the Crab flare in September 2010, to test if their properties are affected by the flare.

\section{Optical Analysis}

\subsection{Relative photometry of HST exposures}
\begin{figure}
\includegraphics[width=65mm]{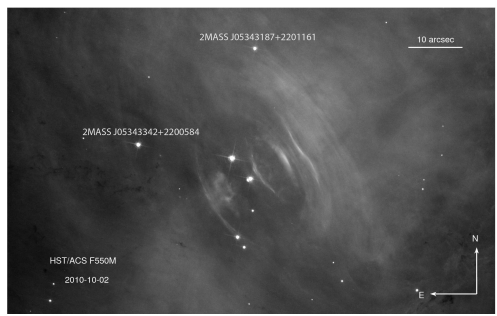}
\caption{\label{reference_stars}HST exposure from October 2th 2010 with reference stars}
\end{figure}

\begin{figure}
\includegraphics[width=65mm]{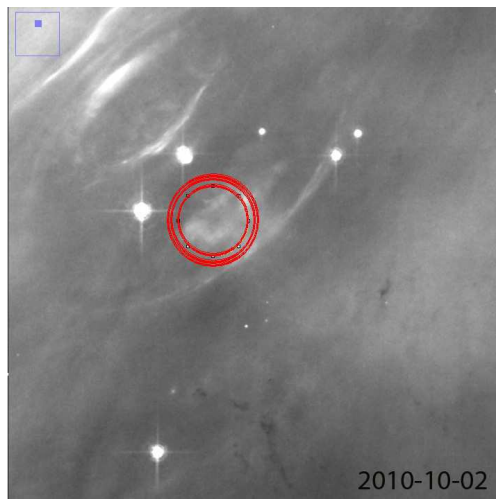}
\caption{\label{relative_brightness}Measurement of relative brightness of the anvil region with ImageJ}
\end{figure}

We examined three exposures from the archive of the Hubble Space Telescope\footnote{http://archive.eso.org/archive/hst/} made in combination with the ACS camera and the F550M filter (Table~\ref{exposures}).
Since the emission from the Crab flare in September 2010 was identified to come from the anvil region located near the pulsar, its relative brightness 
in pixel rates was determined in each exposure 
together with the relative brightness of two environmental stars known as 2MASS J05343342+2200584 and 2MASS 05343187+2201161 (Figure~\ref{reference_stars}). 
The measurement was carried out with the image processing program ImageJ\footnote{http://rsbweb.nih.gov/ij/}. 
With its aperture photometry tool 
the source minus sky brightness was determined of both reference stars as well as the anvil region. In each case the pixel rate was measured within three circles of different radii to 
estimate the brightness of the target and the immediate background (Figure~\ref{relative_brightness}).\\
Comparing the source minus sky brightness of the anvil region in units of each one of the reference stars (units of 2MASS J05343342+2200584 expressed as $\zeta$ and 
in the case of 2MASS 05343187+2201161 as
$\eta$) resulted in a higher increase in the exposure from 2.10.10 regarding both reference stars than in the other two exposures (compare Table~\ref{source_minus_sky_brightness})
which verifies an increased optical emission from the anvil region shortly after the detected flare.

%


\begin{table}[t]
\begin{center}
\caption{Examined HST exposures}
\begin{tabular}{|l|c|c|c|}
\hline \textbf{Date} & \textbf{Exposure Time [s]} & \textbf{Filter} &
\textbf{Instrument}
\\
\hline 2003-08-08 & 2200 & F550M & ACS \\
\hline 2005-11-25 & 2300 & F550M & ACS \\
\hline 2010-10-02 & 2000 & F550M & ACS \\
\hline
\end{tabular}
\label{exposures}
\end{center}
\end{table}

\begin{table}[t]
\begin{center}
\caption{Source minus sky brightness of both reference stars and the anvil region resulting from measurements with ImageJ}
\begin{tabular}{|l|c|c|c|c|c|c|}
\hline \textbf{Date} & \textbf{2MASS*584} & \textbf{2MASS*161} &  \textbf{Anvil region} & \textbf{$\zeta$} & \textbf{$\eta$} 
\\
\hline 2003 & 1293568$\pm$1394 & 506084$\pm$398 & 290420$\pm$55932 & 4.45 & 1.74 \\
\hline 2005 & 874301$\pm$1226 & 397218$\pm$586 & 217727$\pm$4621 & 4.01 & 1.82 \\
\hline 2010 & 434068$\pm$756 & 212004$\pm$470 & 92070$\pm$5271 & 4.71 & 2.3 \\
\hline
\end{tabular}
\label{source_minus_sky_brightness}
\end{center}
\end{table}


\subsection{Synchrotron Model}
In the next part of the optical analysis a synchrotron emission model was determined for the electrons and positrons emerging from the anvil region which was approximated as
a spherical region of 3.5 acrseconds angular radius in a distance of 5 arcseconds from the Crab pulsar as deduced from a Chandra ACIS exposure of the Crab Nebula\footnote{http://chandra.harvard.edu/photo/2008/crab/}.\\
Under the assumption of the Crab pulsar being a constant emitter of electrons and positrons which get their high speed and energy by the induced electric field, 
we set their distribution in the form of a power law:\\

\begin{center}
n(E)dE=KE$^{-q}$dE  
\end{center}

(E: Energy of electrons in Joule, n: Number density of electrons per unit interval in m$^{-3}$ J$^{-1}$, q: Dimensionless power-law-index, K: Normalisation 
coefficient in units of m$^{-3}$ J$^{q-1}$.\\
Since the electrons and positrons emitted by the Crab Nebula have different energies, the values of q and K change. 
Referring to \cite{moroz_1960} several values for the Normalisation coefficient were determined under the assumption of the total number of emitted electrons and positrons 
as 2 x 10$^{-8}$ cm$^{-3}$ which are accelerated in the nebula by a magnetic field with B $\approx$ 10$^{-7}$T (\cite{moroz_1960}, \cite{oort_1956}).
Due to their velocities they are urged on circular orbits by Lorentz force and emit electromagnetic radiation by gyration.
To determine the power of a large number of electrons and positrons gyrating in the magnetic field of the Crab Nebula, the power of a single relativistically gyrating electron
was multiplied with a number of electrons with the energy E which resulted in the total emitted power per unit volume and frequency (\cite{longair_2002}, \cite{rybicki_1979}):

\begin{eqnarray*}
p(f) = \int^{\infty}_{0} P(f,\gamma(E)) KE^{-q} dE
\end{eqnarray*}

In connection with our study of HST exposures we filtered out optical wavelengths via choosing the appropriate border values for the Lorentz factor $\gamma$.
The Normalisation coefficient K was determined in a separate computational analysis.

\begin{figure}
\includegraphics[width=65mm]{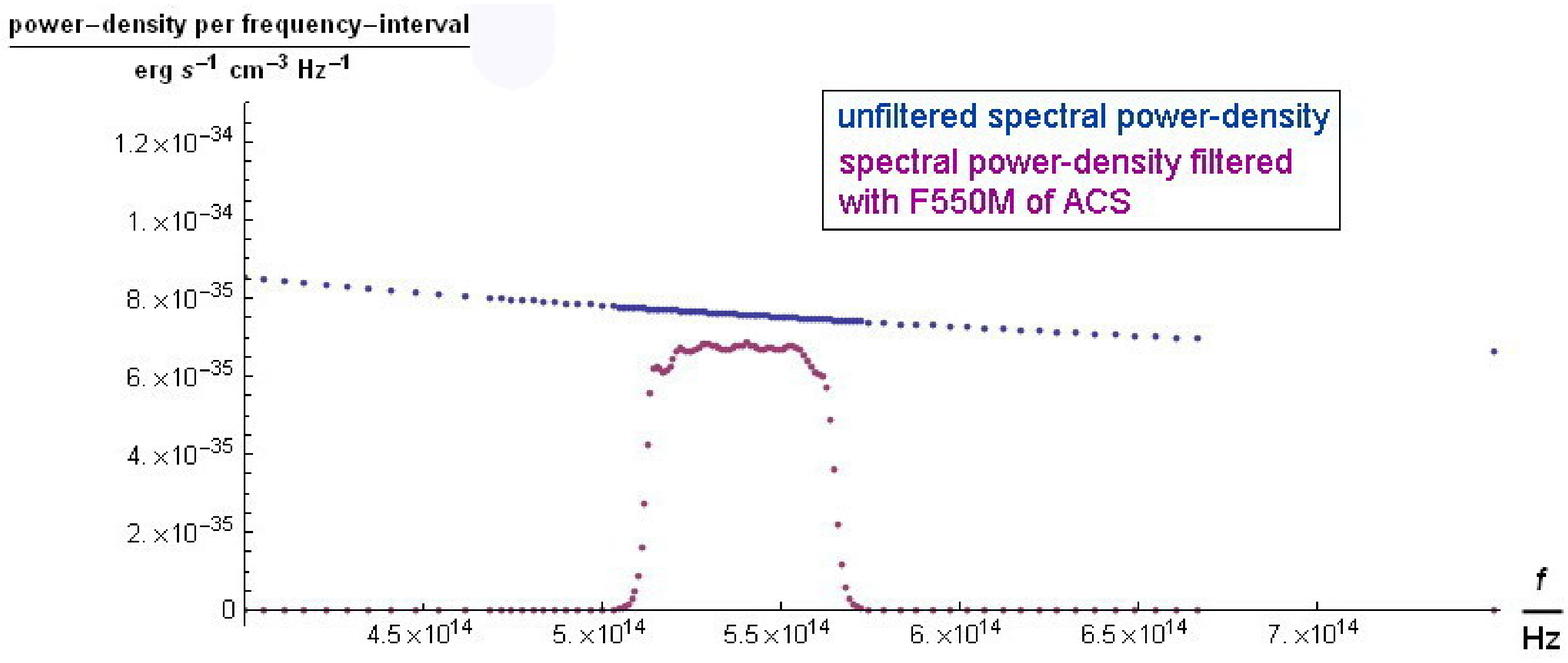}
\caption{\label{result}Spectral Power density of an ensemble of electrons and positrons emitted from the Crab pulsar into the nebula (blue) and with the contribution of the F550M filter (magenta)}
\end{figure}

Since all HST exposures used in this study were made through the F550M filter, we included the respective transmittance to the determined energy distribution (see Figure~\ref{result}).
The total synchrotron emission power resulting from all electromagnetic radiation was determined as P$_{em}$ $\approx$ 10$^{33}$ erg s$^{-1}$ while including the transmittance of
the filter resulted in P$_{filter}$ $\approx$ 10$^{31}$ erg s$^{-1}$. The optical emission power resulting from this analysis amounts to P$_{em}$ $\approx$ 10$^{32}$ erg s$^{-1}$.
It is generally expected that electrons with a power law distribution inherit an energy of $\sim$ 10$^{42}$ erg.\\
Thus the increase of optical emission of the anvil region corresponds with the estimation gotten from the HST exposure and could be calculated with the established synchrotron emission model.

\section{Radio Analysis}

\begin{figure}
\includegraphics[width=65mm]{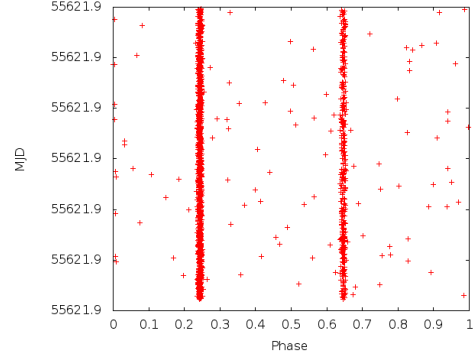}
\caption{\label{GPs_7sigma} Extracted radio GPs from the same data set mainly aligned with P1 and P2}
\label{l2ea4-f1}
\end{figure}

All radio observations used for this study were carried out with the Westerbork Synthesis Radio Telescope (WSRT) at a frequency of 1.38 GHz.
The data sets were coherently dedispersed with the open source pulsar data processing software package
DSPSR\footnote{http://dspsr.sourceforge.net/} and the dispersion measure provided by Jodrell Bank\footnote{http://www.jb.man.ac.uk/$\sim$pulsar/crab.htm}.
As mentioned GPs are apparently phase-bounded, that is they are aligned with regular pulses. To filter them out from each data set, the rms treshold value was set 
to 7$\sigma$ which resulted in an amount of over 1500 GPs from 1.5 h of observations (Figure~\ref{GPs_7sigma}).\\
The selected GPs were flux calibrated with the modified radiometer equation (\cite{lorimer_kramer_handbook}).\\
In the following part of the analysis several characteristics of GPs are examined.
At first the separation times between occuring GPs are determined. According to \cite{karuppusamy_2010} the rate of GPs at the phases of P1 and P2 is one GPs every 0.803 seconds.
A change in the separation times between occuring GPs could prove a change in the GP rate and thus differences in the rotation period of the pulsar.\\
Radio GPs contain intensities which follow a power law distribution (\cite{argyle_1972}) in contrast with the intensities of regular pulses which show a Gaussian distribution 
(\cite{hesse_1974}). In our analysis we examine the brightest GPs resulting from the extraction of a 7$\sigma$ treshold to search for a possible decrease.\\
Since the shortest GPs have been observed to contain the highest flux densities (\cite{sallmen_1999}, \cite{hankins_2003}), we examine the width distribution for all extracted GPs in search for a variation.
The results of this analysis will be shown in a forthcoming publication.

\section{Summary}
Within the framework of this report we present an extensive analysis of radio GPs from the Crab pulsar. 
The initial point given by the increased optical emission power from the anvil region resulting from an optical analysis, gives rise to the question of the 
storage location standing behind these flares.\\
GPs represent short, energetic radio pulses which are thought to develop near the light cylinder through reconnection events. The energy for the latter is located in the magnetic
field of the pulsar. To test if the energy for these reconnection events is affected by the Crab Nebula flare from September 2010, we examine radio GPs observed
after the flare with regard to their rate, pulse width and pulse intensity.\\

\section{Acknowledgement}
The Westerbork Synthesis Radio Telescope is operated by the ASTRON (Netherlands Institute for Radio Astronomy) with support from the Netherlands Foundation for Scientific Research (NWO).\\
We would like to thank Roy Smith, Gyula Jozsa, Gemma Janssen and the whole Westerbork crew for their essential support in carrying out and processing the radio observations.

\end{document}